\documentclass[prl,twocolumn,showpacs,superscriptaddress]{revtex4}
\usepackage[dvips]{graphicx}
\usepackage{dcolumn}
\usepackage{bm}
\usepackage{amssymb}
\usepackage{textcomp}

\DeclareGraphicsExtensions{eps}

\begin{document}
\preprint{APS/123-QED}
\title{Quantum magnetism in the paratacamite family: towards an ideal kagom\'{e} lattice}
\author{P. Mendels}
\author{F. Bert}
\affiliation{%
Laboratoire de Physique des Solides, UMR CNRS 8502, Universit\'{e}
Paris-Sud, 91405 Orsay, France.}%
\author{M.A. de Vries}
\affiliation{%
School of Chemistry and CSEC, The University of Edinburgh, Edinburgh, EH9 3JZ, UK  }%
\author{A. Olariu}
\affiliation{%
Laboratoire de Physique des Solides, UMR CNRS 8502, Universit\'{e}
Paris-Sud, 91405 Orsay, France.}%
\author{A. Harrison}
\affiliation{%
School of Chemistry and CSEC, The University of Edinburgh, Edinburgh, EH9 3JZ, UK  }%
\author {F. Duc}
\affiliation{%
Centre d'\'Elaboration des Mat\'eriaux et d'\'Etudes
Structurales, CNRS UPR 8011, 31055 Toulouse, France}%
\author {J.C. Trombe}
\affiliation{%
Centre d'\'Elaboration des Mat\'eriaux et d'\'Etudes
Structurales, CNRS UPR 8011, 31055 Toulouse, France}%
\author {J.S. Lord}
\affiliation{%
ISIS Facility, Rutherford Appleton Laboratory, Chilton, Didcot, Oxon, OX11, OQX, United Kingdom}%
\author {A. Amato}
\affiliation{%
Laboratory for Muon Spin Spectroscopy, Paul Scherrer Institute, CH-5232 Villigen PSI, Switzerland}%
\author {C. Baines}
\affiliation{%
Laboratory for Muon Spin Spectroscopy, Paul Scherrer Institute, CH-5232 Villigen PSI, Switzerland}%

\date{October 19, 2006}

\begin{abstract}
We report $\mu$SR measurements on the $S$=1/2 (Cu$^{2+}$)
paratacamite Zn$_x$Cu$_{4-x}$(OH)$_6$Cl$_2$ family. Despite a Weiss
temperature of $\sim$~-~300~K, the $x=1$ compound is found to have
no transition to a magnetic frozen state down to 50~mK as
theoretically expected for the kagom\'{e} Heisenberg
antiferromagnet. We find that the limit between a dynamical and a
partly frozen ground state occurs around $x=0.5$. For $x=1$, we
discuss the relevance to a singlet picture.
\end{abstract}

\maketitle

The idea of destabilizing a N\'{e}el state in favor of a spin liquid
state by means of geometrical frustration, was proposed long ago by
Anderson in the context of triangular S=1/2 antiferromagnets
(AF)~\cite{Anderson73}. The underlying concept of a resonating
valence bond (RVB) state, built from a macroscopic number of
singlets, has proven since then to be a very rich theoretical
playground and has indeed been advocated in the important context of
hole-doped quantum AF such as High-$T_c$ cuprates, spin ladders,
etc. The simpler and fundamental case of geometrically frustrated AF
insulators has been deeply revisited for the last 15 years. On a
theoretical ground, there is now a consensus that a RVB state should
be the ground state for $S=1/2$ kagom\'{e}~\cite{Lecheminant, Mila}
or pyrochlore~\cite{Canals} Heisenberg AF (HAF), but not for the
more connected triangular lattice. The most striking results from
exact diagonalizations of the $S=1/2$ kagom\'{e} HAF are the fairly
small gap value $< J/20$ and the existence of a continuum of low
lying singlet excitations, between the ground state and the first
excited triplet~\cite{Lecheminant,Misguich}.

Non-conventional dynamics were experimentally revealed in various
corner sharing AF~\cite{UemuraSCGO,Gardner,Bono}, which yielded the
best support to a resonating-like ground state but no satisfactory
experimental realization has been yet available to check in detail
the $T=0$ theoretical predictions for $S=1/2$ systems, e.g. the
existence of a singlet-triplet gap. Indeed, in a context where all
perturbations such as single ion, exchange anisotropy or spinless
defects, are relevant to induce various marginal orderings, no
material has been fulfilling the condition of a $S=1/2$ perfect
kagom\'{e} HAF lattice. In 
[Cu$_3$(titmb)$_2$(OCOCH$_3$)$_6$].H$_2$O 2$^{nd}$ and 3$^{rd}$
neighbor interactions play a dominant role~\cite{Katsumata,
Domenge}. Volborthite is maybe the first $S=1/2$ kagom\'{e}
lattice~\cite{Hiroi} although one suspects spatially anisotropic nn
interactions; it undergoes a transition to a peculiar spinglass
state around 1~K and has persisting $T\rightarrow 0$
dynamics~\cite{UemuraVolb,BertVolb}. As for 3D structures, rare
earth pyrochlores are up to now recognized as perfect corner-sharing
lattices but most of their fascinating properties are governed by
the delicate balance between magnetocrystalline anisotropy, exchange
and, due to a large $S$ value, dipolar interactions. To our
knowledge, clinoatacamite Cu$_4$(OH)$_6$Cl$_2$ would be the first
example approaching a $S=1/2$ (Cu$^{2+}$) Heisenberg pyrochlore
lattice but it has a distorted lattice and possibly a ferromagnetic
coupling along the $c$-axis. A recent $\mu$SR study clearly revealed
two magnetic transitions, one around 18~K and the other one around
6~K to a state where a weak ferromagnetic moment appears and where
dynamics persist down to $T=0$~\cite{clino_muons}.

Quite recently, the $x=1$ compound of the Zn-paratacamite family
Zn$_x$Cu$_{4-x}$(OH)$_6$Cl$_2$ has been revealed as a "structurally
perfect $S=1/2$ kagom\'{e} antiferromagnet"~\cite{Shores:05}. It can
be viewed as a double variant of the parent compound clinoatacamite
($x=0$); first, the symmetry relaxes from monoclinic to rhombohedral
($R\overline{3}m$) around $x=0.33$, leading to a perfect kagom\'{e}
lattice in the $a$-$b$ plane; then, in the $c$- elongated $x>0.33$
pyrochlore structure, the magnetic bridge along $c$-axis between
$a$-$b$ kagom\'{e} planes is progresively suppressed by replacing
the apical Cu$^{2+}$ by a non-magnetic Zn$^{2+}$. At the composition
$x=1$, the $S=1/2$ Cu$^{2+}$ ions lie at the vertices of a perfect
kagom\'{e} lattice. Whereas macroscopic susceptibility indicates
that the Cu-Cu antiferromagnetic interactions yield a Weiss
temperature $\theta_{\rm{CW}}\sim~- 300$~K, no signature of a
transition was found down to 1.8~K.

In this Letter, we present the first sub-Kelvin magnetic study of
ZnCu$_3$(OH)$_6$Cl$_2$ using $\mu$SR, which is very sensitive to any
freezing and also perfectly suited to track persisting $T=0$
dynamics. We find neither evidence of a magnetic transition nor a
drastic slowing down of the dynamics down to 50~mK, a very original
behaviour in comparison to the other existing geometrically
frustrated magnets. We also track the crossover from a
paramagnetic-like ground state to a $T=0$ magnetic frozen state from
$x=1$ to $x=0$.

$0.15\leq x\leq 1.00$ pure samples where prepared using the
hydrothermal method as described in \cite{Shores:05}. To make the
deuterated analog of the $x=1$ phase (D-sample) the entire reaction
was carried out in a nitrogen filled glovebag, deuterated Copper
Carbonate was prepared and 99\% D$_2$O was used throughout. The
purity of the samples and Cu to Zn ratio were verified ($\approx
\pm1\%$) with powder x-ray diffraction and inductively coupled
plasma - atomic emission spectroscopy (ICP-AES).
 SQUID tests were performed and gave results in perfect agreement
with those of ref.~\cite{Shores:05}. For $x=1$, we find
$\theta_{\rm{CW}}=-300(20)$~K and we detect a small Curie term which
would correspond to a maximum of 6\% $S$=1/2 free spins out of the
total Cu amount. In all our $x\leq 1$ samples a ferromagnetic frozen
component appeared which weight was found negligible for $x=1$ as
in~\cite{Shores:05}, and increased drastically for $x<0.5$.

\begin{figure}
\includegraphics[width=8cm]{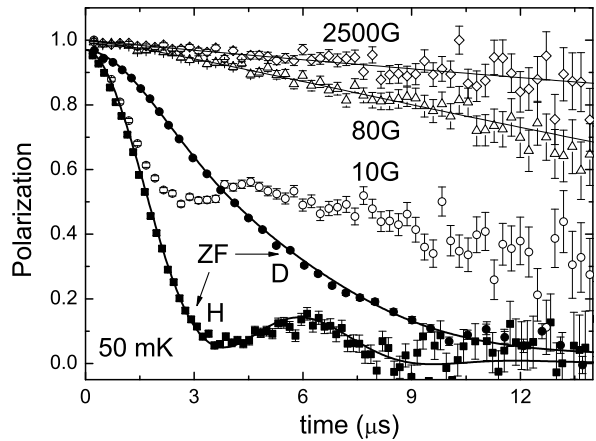}
\caption{$\mu^{+}$ polarization in $x=1$ compound, either in zero
field (H- and - D samples) and under an applied field (H-sample). No
static internal field other than of nuclear origin is detected and a
weak dynamical relaxation is observed.}
\end{figure}

$\mu$SR measurements were performed on $x=0.15$~-~1.0 Zn contents,
in zero field (ZF), longitudinal applied field (LF) or weak
transverse field (wTF) - with respect to the muon initial
polarization. Due to their positive charge, implanted muons come at
rest in well defined sites where their electrostatic energy is
minimized, \emph{i.e.} nearby Cl$^{-}$ and OH$^{-}$. In a
paramagnetic state, electronic moments fluctuate fast ($\sim
10^{-12}s)$ on the $\mu$SR time scale and the field sensed by the
muons has only a nuclear origin with a typical value of a few G.

We first focus on the low-$T$ state for the "perfect kagom\'{e}"
composition, $x=1$ ($T=50$~mK, fig.~1). Damped ZF oscillations,
corresponding to a few G field are evident for the H-sample and were
observed at all $T$, but disappear for the D-sample. They are
therefore clearly associated with the H-nucleus. This is evidence
for the formation of a OH-$\mu$ complex where the muon binds to a
hydroxyl group in the H-sample~\cite{Schenck}. The signature of such
a complex lies in oscillations of the polarization. The
characteristic frequency $\omega_{\rm OH}$ of the time evolution of
the polarization ${\rm P_{OH}}(t)$ can be calculated by modeling the
dipolar interaction between the proton and the muon
moments~\cite{Lord}. In the D-sample, the frequency $\omega_{\rm
OD}$ is too small (nuclear moments ratio $\mu_D/\mu_H=0.153$) and
the oscillating pattern escapes our 0-16~$\mu$s time window. As
expected in a static case, a longitudinal field $\sim$ a few 10~G,
much larger than nuclear dipole fields, decouples nearly completely
the relaxation, except a small relaxing tail which is found
identical in 80~G and 10~G data. The latter can be safely attributed
to a dynamical relaxation, due to fluctuations of the electronic
moments, which only starts decoupling for much larger LF such as
2500~G. In addition, we do not find any change of the oscillations
with $T$ which evidences that there is no static field other than
the dipolar field from nuclei. We can qualitatively conclude that
(i) \emph{no freezing} (ii) \emph{no drastic slowing down} such as
observed in kagom\'{e}-like compounds~\cite{UemuraSCGO, Bono,
UemuraVolb,BertVolb} occur even at 50~mK.

In order to refine this qualitative statement, the ZF polarization
for the H-sample was fitted to the sum of the two expected
contributions,
\begin{eqnarray}
     P_z(t)=[{\rm P_{OH}}(t)e^{-\frac{\displaystyle(\gamma_\mu \Delta_{\rm OH}t)^2}{\displaystyle{2}}}+{\rm KT_{Cl}}(t)]
    e^{(-\lambda t)^\alpha}
\end{eqnarray}
The Gaussian damping $\Delta_{\rm OH}$ originates from nuclei
surrounding the OH-$\mu$ complex. ${\rm KT_{Cl}}(t)$ is the well
established Kubo-Toyabe function, Gaussian at early times used here
for the Cl$^{-}$ site ($\sim$~10~-~20\%) where the relaxation is due
to surrounding Cu and H nuclei. In addition, a small dynamical
relaxation was allowed through the stretched exponential overall
multiplying factor~\cite{relax}. All the static values used in the
fits were fixed by a high statistics run at 1.5~K and found
identical for higher $T$ values and other $x$ values - provided the
samples are in a paramagnetic state. From the value of $\omega_{\rm
OH}$, we find $H_{\rm OH}=7.6(2)$~G, and the field distributions are
respectively $\Delta_{\rm OH}=2.2(3)$~G and $\Delta_{Cl}= 1.7(5)$~G
for OH$^-$ and Cl$^-$ sites, typical of dipolar fields induced by
nearby H, Cu and Cl nuclei. The D-sample data can be consistently
analyzed using $H_{\rm OD}=H_{\rm OH}\times \frac{\mu_D}{\mu_H}$ and
similar $\Delta$ values. At 50~mK we found identical $H_{\rm OH}$
values and only the dynamical relaxation was found to slightly
increase.

The absence of variation of the frequency of the oscillation with
$T$ on the H-sample demonstrates that although
$\theta_{\rm{CW}}\sim-300$~K, the sample does not freeze into a
magnetic state down to the lowest $T$ of 50~mK. Indeed, in the case
of a magnetic freezing, the oscillation due to the OH-$\mu$ complex
would be swamped by the relaxation due to electronic moments. An
upper value of the electronic field of $\sim$ 0.3~G corresponds to
the error bars of our fits. From the comparison with the $\sim$500~G
field found at the muon site in Cu$_4$(OH)$_6$Cl$_2$, one obtains an
upper bound of 6~10$^{-4}$$\mu_B$ for the Cu$^{2+}$ frozen moment.

\begin{figure}
\includegraphics[width=8.5cm]{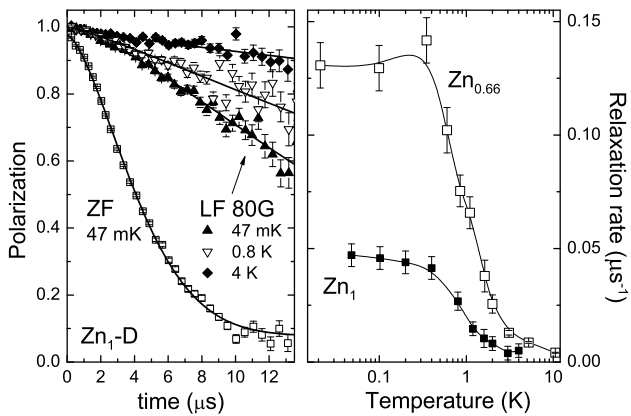}
\caption{Left: time-evolution of the polarization for $x=1$ D-sample
in ZF and LF. The slowing down observed below 1~K is more pronounced
for $x=0.66$ than for $x=1$ (right).}
\end{figure}

We now turn to the discussion of the small dynamical effects
observed on the long time tail of the muon-polarization. We tracked
its $T$-evolution by applying a 80~G longitudinal field, large
enough to decouple the static nuclear fields. In the left panel of
fig.~2, one can directly notice on the D-sample raw data a small
increase of the muon-relaxation between 1.5 and 50~mK. The values of
$\lambda$ extracted from stretched exponential fits are plotted in
the right panel of fig.~2. The modest increase of $\lambda$ is found
to occur around 1~K. The stretched exponent $\alpha$ was also found
to increase from 1 to 1.5 in this $T$ range which reminds of the
unconventional undecouplable Gaussian ($\alpha$=2) observed in a
broad variety of geometrically frustrated magnets~\cite{UemuraSCGO,
Bono}.

In the simplest model of a single time relaxation process, $\lambda$
is related to the fluctuation frequency $\nu$ and the  fluctuating
electronic field $H_{fluct}$ by $\displaystyle{\lambda=2
\gamma_{\mu}^2 H_{fluct}^2 \nu / (\nu^2 + \gamma_{\mu}^2
H_{LF}^2)}$. From fits of the 80~G and 2500~G data at 50~mK (see
Fig.~1), we get $\nu \sim 150$~MHz and $H_{fluct} \sim 18$~G. This
fluctuating field is much smaller than the static field detected in
the frozen phase of the $x=0$ compound ($\sim 500$~G) or the mere
estimate of the dipolar field created by one Cu spin at the oxygen
site ($\sim 2300$~G). Since muons are only sensitive to slowly
fluctuating spins which are not paired in a singlet state, we can
safely conclude that most of the Cu spins appear to be inactive at
50~mK, \emph{i.e.} either in a singlet state, or too fast
fluctuating to be sensed by the muon. Slowly fluctuating defects are
therefore responsible for the small relaxation we observe.

\begin{figure}
\includegraphics[width=9cm]{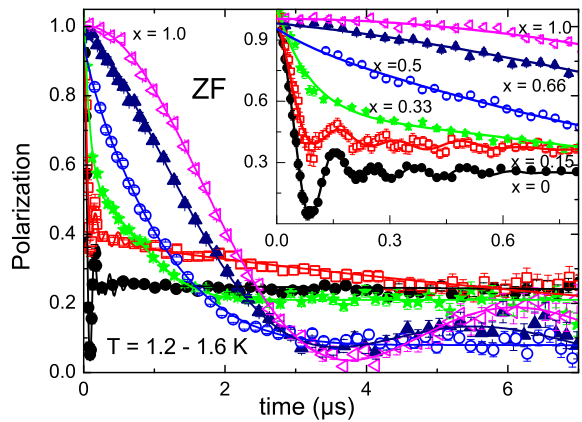}
\caption{$\mu^{+}$ polarization~\cite{Bkgd} plotted versus time for
$x=0,0.15,0.33,0.5,0.66,1.0$. Lines are for fits described in the
text. The data for $x=0$ is from~\cite{Mendelsunpublished}.}
\end{figure}

In order to make the bridge between the fully fluctuating ground
state observed for $x=1$ and the freezing observed for $x=0$, we
performed a detailed $T$-study of the evolution of the magnetic
properties between these two compositions. In fig.~3, we plot the
typical polarizations obtained at $T\lesssim 1.6$~K, representative
of the ground state. The evolution is best followed starting from
$x=0$ where the freezing of electronic moments corresponds to
well-defined spontaneous oscillations of the polarization associated
with the precession of the $\mu^+$ in internal fields $\sim
500$~G~\cite{Mendelsunpublished} (fig.~3, inset). For $x$=0.15, the
oscillations become less evident and a substantial slowly relaxing
component appears. The weight of the latter keeps increasing up to
$x=0.5$, at the expenses of the fast front end (fig.3, main panel).
The $x=0.66$ sample was found to have a dominant dynamical character
down to 50~mK which we probed through LF measurements such as
described for $x=1$. For this composition, the slowly oscillating
pattern of the OH-$\mu$ complex is also partially recovered. The
slowing down of the dynamics, more pronounced than for $x=1$ is
clearly identified around 1~K (fig.~2, right panel).

For $x\leq0.5$, the ZF data was fitted to a superposition of a fast
relaxing ($x\geq0.33$) or oscillating ($x\leq0.15$) component
corresponding to the frozen fraction, and a paramagnetic term given
by Eq.~1, where all the static fields were kept at the values found
at high $T$ consistent with those for $x=1$ (fig.~3). The
$T$-variation of the frozen fraction, determined through 20~G wTF
measurements, is plotted in fig.~4. For $x=0.15$, two transitions
clearly occur around 18 and 6~K but, at variance with the $x=0$
sample, correspond to different frozen fractions. For the $x=0.33$
and 0.5 sample, the 18~K transition disappears, which therefore
rather seems typical of the monoclinic phase ($x\leq 0.33$). A very
broad transition is still observed around the same $T\sim 6$~K and
with very different frozen fractions. The change in the shape of the
transition curves indicates that no macroscopic phase segregation
occurs but rather that for $x\leq 0.5-0.6$, islands freezing around
6~K coexist with paramagnetic ones which have slower dynamics than
for $x=1$. The low-$T$-fractions are represented in the inset versus
$x$ and clearly indicate a change of behavior from partially static
to dominantly dynamical behavior around $x$=0.5-0.6. This certainly
relates to the $x=0.5$ percolation threshold between a limit of
decoupled kagom\'{e} planes and a case where all Cu$^{2+}$ on a
triangle belong to at least one fully occupied (elongated)
pyrochlore pattern. Above $x=0.5-0.6$, the dynamics is likely
inhomogeneous due to random local variations in Cu occupancy of the
Zn site. The extent of this region signals, in turn, that the
physics of the kagom\'{e} planes is only little perturbed through
Cu-Cu interactions along the $c$-axis.

\begin{figure}
\includegraphics[width=9cm]{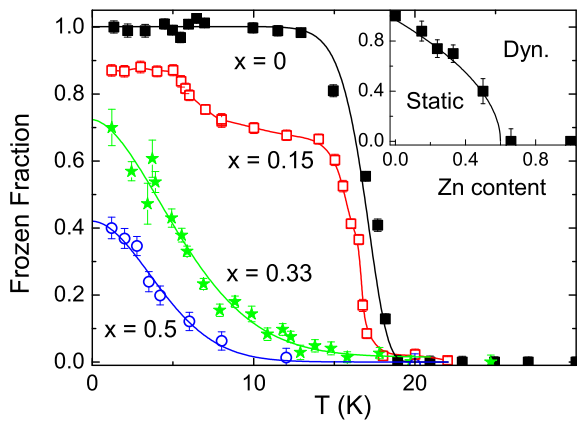}
\caption{$T$-variation of the frozen fraction for all $x$ where a
magnetic transition is observed. Inset: low-$T$ fraction versus $x$.
Symbols used here are the same as in fig.~3. Lines are guides to the
eyes}
\end{figure}

ZnCu$_3$(OH)$_6$Cl$_2$ ($x=1$) represents the first example of a
kagom\'{e} system without any 3D freezing down to temperatures as
low as 1.6~10$^{-4} J$. Its low-$T$ behavior differs widely from the
cases of kagom\'{e} bilayers of the long studied magnetoplumbite
family~\cite{SCGORamirez} or of volborthite. We find no marginal
spin glass state even for one order of magnitude smaller $T/J$ range
and the low-$T$ relaxation is much smaller than the unconventional
one observed in these spin glass-like
phases~\cite{UemuraSCGO,Bono,UemuraVolb}. The explanation for this
cannot be related to the absence of defects since this system also
displays a low-$T$ Curie tail in the SQUID data
($\sim$~few~\%/f.u.). This statement certainly needs to be refined
by a better understanding of the nature of these defects in Zn-
paratacamite.

Two scenarios could be at work to explain the low-$T$ dynamical
properties. The first one traces back to the RVB initial picture or
as well to any singlet ground state~\cite{Misguich}. If the
singlet-triplet gap is large as compared to the $T$ range of the
data presented here, one should not detect any dynamics through the
muon, which senses magnetic fields only. The remaining dynamics
could be due to singlets which are broken by the defects and would
yield either deconfined unpaired spins (sporadic
excitations~\cite{Bono}) as expected in a RVB model or localized
unpaired spins, depending on the nature of the ground state. It is
worth noting that we did not find any drastic change in the $\mu$SR
relaxation up to 150~K. This does not necessarily means the absence
of a paramagnetic to singlet phase transition since in these phases
the dynamics is either too fast for the muon or absent and the
transition might be hard or even impossible to detect through
$\mu$SR. In this scenario, the modest change in the relaxation
between the $x=0.66$ and 1 samples, as compared to other kagom\'{e}
compounds~\cite{Bono,UemuraVolb} could be well explained by an
unchanged concentration of defects in the kagom\'{e} plane since
only the apical Zn is replaced by Cu when $x$ is decreased. The
details of the interactions between defects and apical Cu$^{2+}$
ions might govern the low-$T$ dynamics and could be responsible for
the small change observed in the relaxation rate. The second
scenario relies on the idea that the singlet-triplet gap is smaller
than 50~mK, despite a 300~K $J$-value. The density of excitations
would still be quite high and ZnCu$_3$(OH)$_6$Cl$_2$ would then
reproduce the essential features of a "cooperative paramagnet" such
as observed in Tb$_2$Ti$_2$O$_7$ spin liquid state~\cite{Gardner}
which has no transition down to 70~mK. However, these systems may
only be loosely related since Tb$_2$Ti$_2$O$_7$ with large and
Ising-like spins likely belongs to another class of frustrated
magnets, especially if one refers to the ordered spin-ice behavior
found in the parent Tb$_2$Sn$_2$O$_7$
compound~\cite{Mirebeau,BertPyro}.

In conclusion, Zn-paratacamite is a very promising candidate for the
$S=1/2$ kagom\'{e} lattice model. The issues of the existence of a
gap and of the nature of low energy excitations will certainly be
major avenues for future research in the field of geometrically
frustrated magnets. The explanation of the different dynamical
behaviors and ground states of various kagom\'{e}-like compounds
will certainly represent a second challenge and not only the $x=1.0$
compound is interesting but also the broad range of $x$ values above
0.5-0.6 which provide a way of probing the physics through
perturbative 3D couplings.

P.M. thanks C.~Lhuillier for bringing Zn-paratacamite to his
attention. This work was supported by E.C. FP 6 program, Contract
RII3-CT-2003-505925. MdV acknowledges support from the HFM network
of the ESF.


\begin{thebibliography}{19}
\bibitem{relax}Rigorously, one relaxation rate per site should be
introduced but the fits do not allow a reliable determination.
\bibitem{Bkgd}Polarizations are deduced by
subtracting a background from raw asymmetries and normalizing the
data to 1. They are therefore defined within a 0.05 uncertainty.
\end{thebibliography}
\end{document}